\documentclass[aps,prl,twocolumn,longbibliography,10pt]{revtex4-2}

\usepackage{amsmath}
\usepackage{upgreek}
\usepackage{amssymb}
\usepackage{graphicx}
\usepackage{bbold}
\usepackage[utf8]{inputenc}
\usepackage[T1]{fontenc}
\usepackage{color}
\usepackage{braket}
\usepackage{float}
\usepackage[english]{babel}
\usepackage{hyperref}
\usepackage{cleveref}
\usepackage{tikz}
\usepackage{booktabs}
\usepackage{tabularx}
\usepackage{threeparttable}
\usepackage{siunitx}
\usepackage{array}
\newcommand{\figpanel}[2]{%
  \hyperref[#1]{Fig.~\ref*{#1}(#2)}%
}

\usepackage{caption}
\usepackage{subcaption}
\usepackage{ragged2e}

\DeclareCaptionFont{just}{\justifying}
\DeclareCaptionFont{just}{\justifying}
\begin{document}

\title{Quantum Reservoir Computing for Statistical Classification in a Superconducting Quantum Circuit}

\author{J.\ J.\ Prieto-Garcia}
\author{A. G. del Pozo-Martín}
\author{M. Pino}
\affiliation{University of Salamanca IUFFyM, Salamanca, Spain}

\begin{abstract}
We analyze numerically the performance of Quantum Reservoir Computing (QRC) for statistical and financial problems. We use a reservoir composed of two superconducting islands  coupled via their charge degrees of freedom. The key non-linear elements that provide the reservoir with rich and complex dynamics are the Josephson junctions  that connect each island to the ground. We show that QRC implemented in this circuit can accurately classify complex probability distributions, including those with heavy tails, and identify regimes in correlated time series, such as periods of high volatility generated by standard econometric models.  We find QRC to outperform some of the best classical methods when the amount of information is limited. This demonstrates its potential to be a noise-resilient quantum learning approach capable of tackling real-world problems within currently available superconducting platforms. We further discuss how to improve our QRC algorithm in real superconducting hardware to benefit from a much larger Hilbert space.
\end{abstract}

\maketitle

\section{Introduction}

There are several quantum algorithms that have solid computational advantages with respect to their classical counterparts when they are run in a noiseless and scalable quantum computer\ \cite{benioff1980computer, feynman2018simulating,divincenzo2000physical,preskill2018quantum,shor1994algorithms,grover1996fast,kitaev1995quantum,peruzzo2014variational}. 
Unfortunately, such quantum devices are currently out of reach as quantum computers are still too noisy and error correction has not been scaled up yet\ \cite{google2025quantum,he2025experimental,eickbusch2025demonstration, roffe2019quantum}. As a result, it is a fundamental challenge to search for algorithms that can lead to quantum advantages while remaining compatible with current hardware \cite{arute2019quantum,zhong2020quantum,pan2022solving,childs2010quantum}.  An even greater challenge lies in the search for quantum advantages in real-world computational problems. This is the challenge we address in this work.

One algorithm that can help in the search of quantum advantages is the so-called Quantum Reservoir Computing (QRC). It  employs a complex quantum system, a reservoir, to perform supervised learning tasks on an input signal\ \cite{fujii2017harnessing,fujii2021quantum,Mujal_2021, palacios2024role, garcia2023scalable, llodra2025quantum}. The reservoir quantum dynamics is driven by that signal, while an additional classical layer is used to map the reservoir quantum dynamics to the desired output, similarly to classical  reservoir computing\ \cite{kirby1991context,jaeger2001echo,maass2002real}. In analogy with recurrent neural networks \cite{elman1990finding, hochreiter1997long}, the occupation probabilities of the reservoir’s basis states can be viewed as the activations of its neurons. Due to its analog nature, it is relatively easy to implement QRC as  only the intrinsic quantum system dynamics are needed with a minimal amount of control.  QRC is also robust against a finite amount of noise and it can even benefit from a moderate amount of it \cite{Domingo2023,Sannia2024dissipationas,Fry2023}. Last but not least, QRC is able to analyze correlated time series \cite{nokkala2023online, suzuki2022natural, hamhoum2025multivariate, mujal2023time, kutvonen2020optimizing}, a feature that allows the treatment of, for instance, real-world problems in economics and finance \cite{li2025quantum}. 

\begin{figure}[h!]
    \centering
    \begin{subfigure}{0.48\textwidth}
        \centering
        \includegraphics[width=\linewidth]{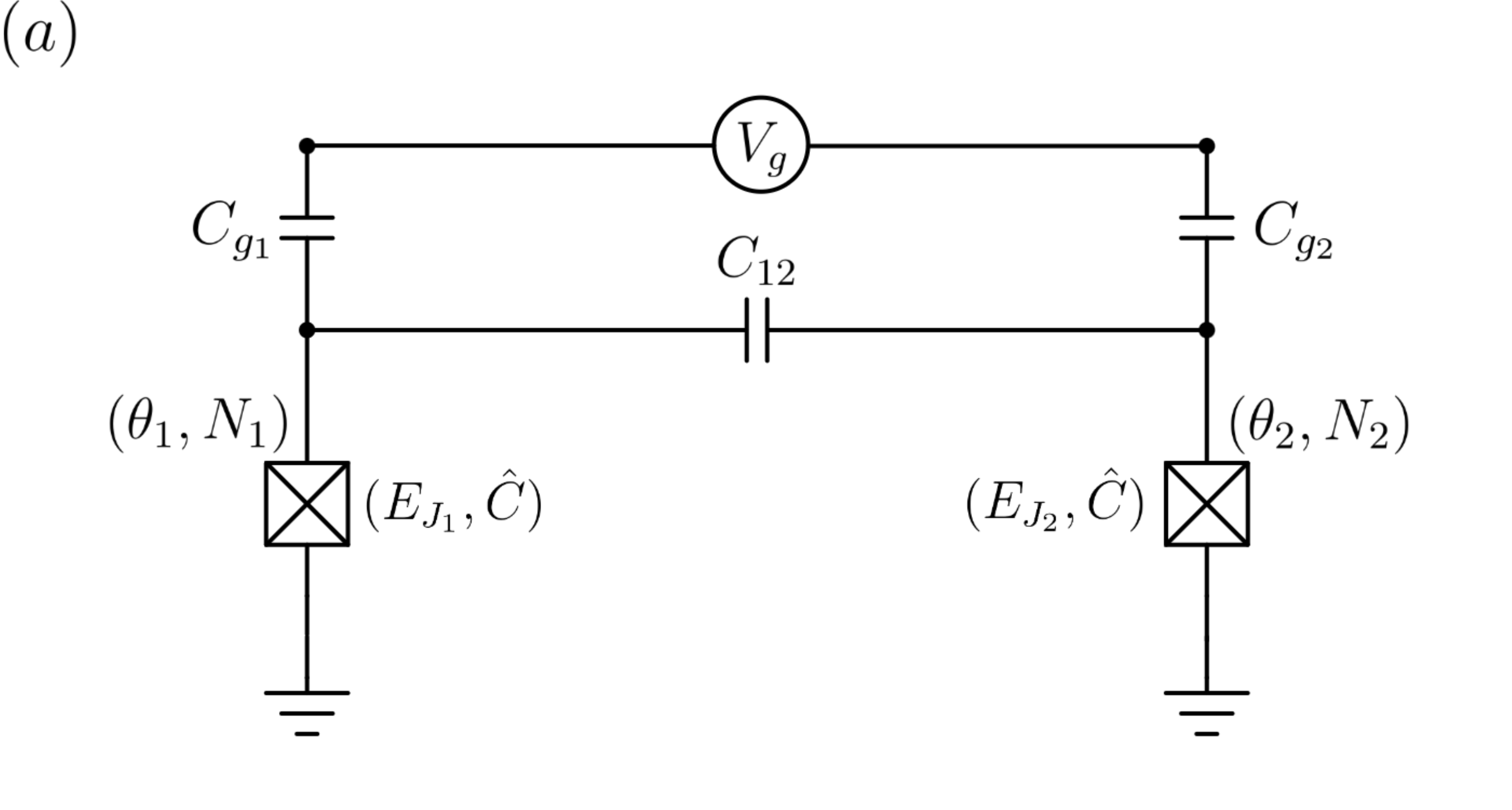}
    \end{subfigure}
    \hfill
    \begin{subfigure}{0.48\textwidth}
        \centering
        \includegraphics[width=\linewidth]{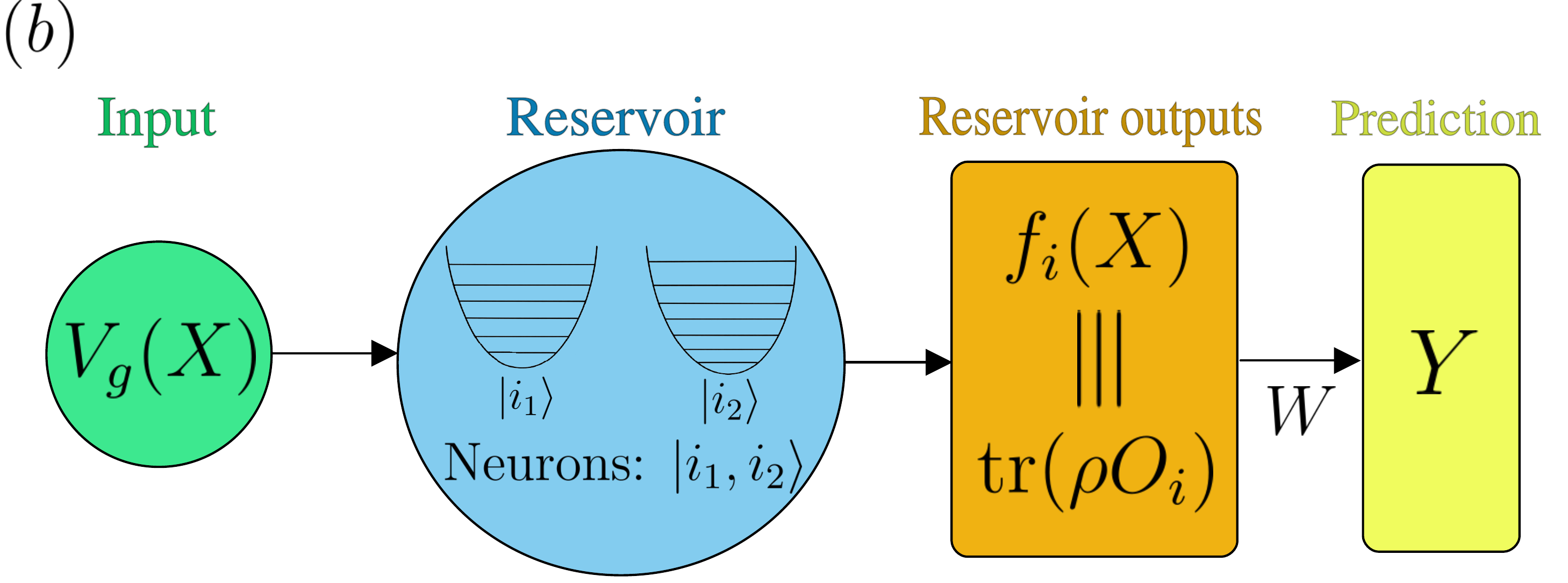}
    \end{subfigure}
    \caption{(a) The quantum reservoir is a superconducting circuit composed of two capacitively coupled  superconducting islands, with phase and charge degrees of freedom $\theta_j,N_j$, which are connected to the ground via Josephson junctions.  Both islands are connected to an external voltage $V_g.$ The coupling capacitance is $C_{12},$ while each gate capacitance is $C_{g_{i}}.$ (b) Representation of the QRC framework. The quantum reservoir (large blue circle) evolves in response to the input (green circle). At specific times, a set of observables $O_i$ is measured from the reservoir and sent to the readout layer. The reservoir outputs (orange square) are subsequently mapped to the reservoir prediction (yellow square) through a set of trained weights (represented by the weight matrix $W$).} 
    \label{fig:circuito_esquema}
\end{figure}

In this work, we analyze QRC for statistical discrimination tasks using as a reservoir a simple superconducting quantum circuit\ \cite{ripoll2022quantum,devoret2013superconducting,krantz2019quantum, pino2015}.  This circuit is composed of two  capacitively coupled superconducting islands which are connected to the ground via a Josephson junction\ \cite{johnson2011quantum, harris2018phase, houck2012chip, georgescu2014quantum}. Those junctions, which operate in the transmon regime, provide the key ingredient to obtain rich reservoir dynamics. Indeed, their inclusion results in the low-energy dynamics of the system being described by the Bose-Hubbard model, which has been widely used to describe interacting bosons\ \cite{dutta2015non, llodra2025quantum, pino2013capturing}, see \figpanel{fig:circuito_esquema}{a}. We emphasize that our QRC algorithm is fully analog,  which is different from previous proposals based on qubit gates\ \cite{suzuki2022natural,li2025quantum}.

We test the performance of our superconducting quantum reservoir on three problems, such as the discrimination of probability distributions and the parameter prediction of fixed distributions. The learning problems we address are difficult, since in some of them we classify distributions that exhibit heavy tails and contain correlations among different input data. Specifically, we first consider the problem of discriminating between two probabilistic families: normal and Laplacian distributions. Second, we tackle the problem of parameter estimation from data generated from Student-t distributions. This requires inferring how heavy the tails of the underlying distribution are, a feature of wide interest in areas such as economics and time forecasting \cite{cont2001empirical,mandelbrot1963variation}. The third problem is to classify volatility regimes in $\text{GARCH}(1,1)$ models, widely used in economic forecasting \cite{engle1982autoregressive, bollerslev1986generalized}. 

We show that our specific realization of QRC provides a general algorithm that is able to compete rather well with some of the best classical counterparts for \emph{each} of the previous tasks. Furthermore, we find that our algorithm improves over classical results when the amount of information from the underlying probability distribution is limited. We will finally argue how our QRC learning approach could be improved by running it in a real superconducting device, where the quantum reservoir can benefit from  a  much larger Hilbert space.

\section{Computational Problems}\label{Sec:problems}

In this section, we explain each of the problems that we aim to solve with QRC. We are interested in solving questions in the realm of statistical inference. We do not restrict to academic problems, but apply QRC to perform statistical inference of data following probability distributions similar to those that appear in several areas of science, such as finance or climate predictions. As we employ QRC as a supervised learning approach to solve this problems, we require two main stages to characterize the QRC efficiency: training and testing. In the following, we carefully describe the three problems that we will deal with. 

\subsection{Normal vs. Laplace distribution inference} \label{sec:problemsGL}
The first task is to infer the probability distribution followed by a list of $T$ data points. We consider data sampled  from either a Normal or a Laplacian distribution. That is,  our algorithms should decide from which distribution a set of data points has been drawn. The density functions of these two distributions are respectively given by:
\begin{align}
    f_{\mathcal{N}}(x;\mu,\sigma)
&= \frac{1}{\sqrt{2\pi \sigma^2}}
\exp\!\left(-\frac{(x-\mu)^2}{2\sigma^2}\right),\label{eq:normal}\\
f_{\mathcal{L}}(x;\mu,b)
&= \frac{1}{2b}
\exp\!\left(-\frac{|x-\mu|}{b}\right),\label{eq:laplace}
\end{align}
where $\mu$ is the mean of each distribution, $\sigma>0$ is the standard deviation of the Normal distribution, and $b>0$ is the scale parameter of the Laplacian distribution.

Note that the goal is to predict the type of distribution regardless of the values of its parameters, and hence to identify the shape of the distribution rather than translations or rescaling. We restrict the possible range of each distribution's parameters to the following: $\mu \in [1,2]$ and $\sigma \text{ or } b\in [0.1,0.7]$. In particular, during the training phase we sweep over these ranges to cover the full parameter space, while in the test phase the parameters are chosen randomly within the same intervals. 

\subsection{Student-t parameter inference}\label{sec:problemsTS}

The second task is to infer some of the parameters of data drawn from a  Student-\textit{t} distribution. Specifically, we will predict the degrees of freedom  $\nu$ for a probability density function given by
\begin{equation} \label{eq:tstudent}
    f_t(x;\mu, \sigma,\nu)
=
\frac{\Gamma\!\left(\frac{\nu+1}{2}\right)}
{\sqrt{\nu\pi}\,\sigma\,\Gamma\!\left(\frac{\nu}{2}\right)}
\left(
1+\frac{1}{\nu}\left(\frac{x-\mu}{\sigma}\right)^2
\right)^{-\frac{\nu+1}{2}},
\end{equation}
where $\mu$ is a location parameter and $\sigma>0$ is a scale parameter. 

The weight of the Student-\textit{t} distribution's tails depends on $\nu$. Indeed,  the smaller the value of $\nu$, the heavier the tail, while for large $\nu$ the distribution tends to a Normal distribution.  Since the goal of this task is to evaluate the ability of the reservoir to identify heavy-tailed distributions, we restrict to the case $\mu=0$ and $\sigma=1$. Furthermore, we focus on predicting the quantity $1/\nu$ rather than $\nu$, since it provides a more linear sensitivity to changes in the tail heaviness. Thus, we sweep $1/\nu$ uniformly over the interval $[1/30,1]$ in the training phase, whereas in the test phase $1/\nu$ is chosen uniformly at random within the same range. 

\subsection{GARCH volatility estimation}\label{sec:problemsGARCH}

The final task is of particular importance, since it exhibits time correlations, which are present in many processes in economics and finance. In numerous real-world series, such as asset returns, periods of large fluctuations remain turbulent for a long time, while calmer periods persist in a similar way. The standard framework to study the so-called volatility clustering phenomenon is the GARCH(1,1) family \cite{engle1982autoregressive, bollerslev1986generalized}, whose dynamics capture how the instantaneous volatility evolves depending on both recent shocks and past volatility levels. Specifically, data at time $t$ are given by:
\begin{align*}
    x_t&=\sigma_t z_t,\\ 
    \sigma_t^2 &= \omega + \alpha x_{t-1}^2 + \beta \sigma_{t-1}^2,
\end{align*}
where $\sigma_t$ is the conditional standard deviation at time $t$ (also referred to as volatility), $\omega$ is related to the long-run variance level, $\alpha$ is a short-term volatility response, $\beta$ is a persistence parameter, and $z_t$ is sampled from a Normal distribution with mean $0$ and variance $1$.

A key structural parameter of these models is the sum $\alpha + \beta$ \cite{tsay2005analysis}, which directly determines the persistence of volatility. Shocks dissipate quickly when that sum is small, whereas values close to one correspond to strong volatility clustering and to heavier tails. Hence, characterizing this persistence is crucial for understanding the underlying volatility regime of the process. Thus, our last task consists of classifying sequences of $T$ data points generated by $\text{GARCH}(1,1)$ models into three distinct persistence bands---low, medium, and high---defined respectively as $\alpha + \beta \in [0.2,0.6], \text{ } \alpha + \beta \in [0.6,0.9], \text{ or } \alpha + \beta \in [0.9,0.99]$. For simplicity, we set $\omega=1$ and train the reservoir with series whose $\alpha \text{ and }\beta$ values are chosen randomly in each interval. The test phase works analogously.

\section{Architecture}

We now introduce the hardware used to deal with the problems presented in the previous section. We analyze a simple superconducting circuit, which will play the role of the reservoir, and derive its low-energy Hamiltonian. We will find this Hamiltonian to be the Bose-Hubbard one of interacting bosons, whose efficiency for QRC has been thoroughly analyzed in Ref.\ \cite{llodra2025quantum}. The derivation presented here will help us to understand meaningful experimental parameters which are realizable in a superconducting circuit and obtain valuable knowledge on which hardware modifications can improve the QRC efficiency. 

We consider a system composed of two coupled superconducting islands connected to ground via equal Josephson junctions. Those islands are capacitively connected to the same external voltage $V_g$, which allows the system to be externally driven, see \figpanel{fig:circuito_esquema}{a}. The i-th island can be described in terms of its number of Cooper pairs and phase,  $N_i$ and $\theta_i$ , respectively.  These dimensionless variables are canonically conjugate and  satisfy $[\theta_i, N_j] = i\delta_{i,j}$ in the quantum description of the system. The  Hamiltonian is
\begin{align} \label{eq:Htotal}
  H=& \sum_{i=1,2}H_{i} +\Delta H,
\end{align}
where the Hamiltonian of each island and the coupling between them are given, respectively, by
\begin{align}
    H_i &=  4E_{C} (N_i-N_{g})^2 - E_{J} \cos(\theta_i)\label{eq:1island}\\
    \Delta H &= TN_1 N_2.\label{eq:coupling}
\end{align}
We denote by $E_J$ and $E_C$ the Josephson and (renormalized) capacitive energies of the junctions. Notice that the capacitive energy  $E_C=e^2/(2C)$ takes into account the renormalized island capacitance $C^{-1}=\left(\hat{C}+C_{12}+C_g\right)/\hat{C}^2$, and not the bare Josephson junction capacitance $\hat{C}.$ The offset charges are denoted by $N_{g}=C_{g}V_g/(2e)$ and depend on the gate capacitor $C_g$ and voltage  $V_g$. The parameter $T$ controls the coupling and depends on  $C_{12}$  and $\hat{C}$. It is given by $T= 4e^2C_{12}/ \hat{C}^2,$ see Ref.\  \cite{hita2022ultrastrong} for more information about the renormalization of the capacitance due to the coupling. We will employ a configuration operating in the regime $E_{J} \gg E_C$.  This is a widely used junction regime, as it corresponds to the regime relevant for transmon qubits, and allows one to map the dynamics of our Hamiltonian in Eq.\ \ref{eq:Htotal} to the Bose-Hubbard model\ \cite{houck2009life, place2021new}.

In order to distill the low-energy model of our circuit, we first simplify the Hamiltonian of each superconducting island $H_i.$ It can be interpreted as a Hamiltonian describing a harmonic oscillator, with frequency $\omega = \sqrt{8E_JE_C}$, plus non-linearities coming from the higher-order terms of the cosine Josephson potential. In the regime in which we are working, $E_J \gg E_C,$ we obtain $E_{J}\gg \omega$, and hence the levels close to the ground state are quite confined around the potential minimum. In this situation, the low-energy states only see a small non-linear correction and it is safe to keep only a few orders in the expansion of the Josephson potential. Considering the cosine expansion up to fourth order in Eq.\ \ref{eq:1island}, we obtain the approximate single-island Hamiltonian
\begin{equation}
    H_{i}  =   4E_{C} N_i^2 +  \dfrac{E_{J}}{2} \theta_i^2 - \dfrac{E_{J}}{4!} \theta_i^4 - 8E_{C} N_g N_i. \label{eq:truncation}
\end{equation}
The first two terms correspond to the ones in a harmonic oscillator with frequency $\omega.$ Hence, using $\theta_i=(2E_C/E_{J})^{1/4} (a_i+a_i^\dagger)$ and $N_i=i(E_{J}/32E_C)^{1/4} (a_i^\dagger - a_i)$, the Hamiltonian takes the form
\begin{equation}
H_i =  \omega a_i^\dagger a_i  - \dfrac{E_{C}}{12} (a_i+a_i^\dagger)^4 -i \varepsilon V_g (a_i-a_i^\dagger).
\end{equation}
where $\varepsilon = 2 e^{-1} E_C C_{g} (E_{J}/2E_C)^{1/4}.$  Since $E_C \ll E_{J}$, it follows that $E_C \ll \omega$, and we can therefore apply the Rotating Wave Approximation \cite{scully1997quantum}, retaining only the number-conserving terms. Thus, the final Hamiltonian is given by
\begin{align}
    H_{i} =  \omega n_i  - \dfrac{E_{C}}{2} n_i(n_i+1) -i \varepsilon V_g (a_i-a_i^\dagger),
\end{align}
where $n_i=a_i^\dagger a_i$ denotes the number operator of oscillator $i$. We now simplify the remaining term $\Delta H$ in Eq.\ \ref{eq:coupling}. To do so,  we assume that  $T$ is sufficiently small  $g=\, T\sqrt{E_{J} /(32 E_C)} \ll \omega,$ so that we can drop counter-rotating terms in the coupling, thus obtaining
\begin{equation}
    \Delta H = g (a_1^\dagger a_2 + a_1 a_2^\dagger).
\end{equation}
Notice that the simplifications made here are accurate as far as the strong coupling regime is avoided. This occurs roughly for couplings up to $g\sim  0.1\omega.$ 

We have finally obtained a Bose-Hubbard Hamiltonian with external drive:
\begin{align} \label{eq:fullmodel}
&H=  \sum_{i=1}^2 \omega n_i - \dfrac{U}{2} n_i(n_i+1)  +  i \varepsilon V_g(a_i-a_i^\dagger)\\
&  \notag + g_{12} (a_1^\dagger a_2 + a_1 a_2^\dagger).
\end{align}
The gate voltage $V_g$ will be used to input data into the system. In QRC $\varepsilon V_g = \varepsilon_0\, x_j$  will be used, with $x_j$ a dimensionless amplitude which contains the data to feed into the system at time $t_j.$ The interacting term is controlled by  $U=E_C,$ which comes from the nonlinearities of the Josephson cosine potential. The phase diagram of this model is an interesting one, composed of superfluid and Mott-insulator phases. It further contains a Berezinsky-Kosterlitz-Thouless critical point and a re-entrant phase-diagram\ \cite{pino2013capturing}. Previous works have reported a good performance of Bose-Hubbard reservoirs in paradigmatic QRC tasks near the phase transition and deep in the superfluid regime\ \cite{llodra2025quantum}. Although these results were obtained for relatively large systems, here we will focus on the case of a Bose-Hubbard Hamiltonian with only a few sites. 

We will use the previous model Eq.\ \ref{eq:fullmodel} to explore QRC numerically.  First of all, as it is impossible to reach arbitrarily large Hilbert space dimensions,  we have used a  cutoff $n_c=5$ in the occupancies of each bosonic mode. This choice still provides a rich dynamical response of the reservoir while reducing considerably the computational cost of the QRC simulations. We further notice that, in the case of increasing this cutoff, our superconducting circuit Fig.\ \ref{fig:circuito_esquema} may not be well captured by the Bose-Hubbard model Eq.\ \ref{eq:fullmodel}. The reason is that higher corrections than the quartic one will need to be included in Eq.\ \ref{eq:truncation}. This is an interesting avenue of research as the resulting model, a pure rotor one, can display larger non-linearities than those considered here\ \cite{pino2016nonergodic}. 

We now specify the parameter choice in Eq.~\eqref{eq:fullmodel}. We employ a coupling $g_{12}\approx 0.05\,\omega$ and anharmonicity $U\approx 1.5\,g_{12}$. For typical superconducting devices, we can then use standard Josephson junctions with $\omega \sim 10\text{ GHz},$ giving a time scale for the system dynamics on the order of $0.1-1\ {\rm ns}$. Although we have treated identical Josephson junctions in our previous analysis, we have implemented the QRC algorithm using slightly different Josephson energies to make the dynamics more complex. Regarding the driving term, it has been intentionally chosen to be relatively large so as to drive transitions toward higher-energy states, as QRC needs to explore such states for better performance, yet sufficiently small to ensure that the population of states near the cutoff remains negligible. Doing so, we guarantee the validity of the truncated Hilbert space description.

Although our theoretical model can be obtained in other quantum hardware, such as optical lattices\ \cite{morchs2006}, we believe that the realization with superconducting circuits can provide several key advantages for QRC. For example, it allows easy variation of model parameters through the fabrication process, individual driving via external voltages or fast operational repetition rates. Similar hardware designs have been used for QRC before \cite{dudas2023, dudas2022coherently, nokkala2023online} but mainly using linear oscillators.

\section{Quantum Reservoir Computing}

We devote this section to describing the principles of QRC.  As shown in \figpanel{fig:circuito_esquema}{b}, it consists of three main components: the input layer, the quantum reservoir, and the readout layer. The input layer receives a time-dependent signal $X\equiv (x_j)$, typically corresponding to data that are not easily separable into distinct classes, and encodes it into a certain physical parameter of the quantum system. In our implementation, the superconducting circuit of the previous section plays the role of the reservoir, so that the input is encoded in the amplitude of the gate voltage $V_g$  in Eq.\ \eqref{eq:fullmodel}. This encoding modulates the system’s internal dynamics, causing the reservoir to evolve through a sequence of configurations that depend on the input signal. Through this process, the quantum reservoir performs a nonlinear transformation projecting the input data into a high-dimensional feature space where they become more easily separable.

At certain times during the reservoir evolution, a set of observables $O_i$ is measured to extract features that characterize the system’s state, represented schematically as an orange square in \figpanel{fig:circuito_esquema}{b}. These measurements are typically chosen so that they correspond to occupation probabilities of selected quantum levels, from which a matrix $F(X)$ is constructed. Then, the objective of the readout layer is to transform the reservoir outputs $F(X)$ to the target, encoded in the matrix $\hat{Y}$, via a weight matrix $W$. To that end, the algorithm is divided into two phases: training and test. 

In the training phase, $W$ is determined by fitting the reservoir outputs to known target values using the Moore–Penrose pseudo-inverse, a standard approach in both classical and quantum reservoir computing~\cite{appeltant2011,brunner2013,dudas2023}:
\begin{equation}
    W = F^{\dagger}(X_\text{training}) \cdot \hat{Y}_\text{training},\label{eq:pseudo}
\end{equation}
where $F^\dagger$ denotes the Moore–Penrose pseudo-inverse of the matrix $F$. In the test phase, the obtained $W$ and $F(X_\text{test})$ are used to compute the predicted values $Y_\text{test}$, which are compared to the test target $\hat{Y}_\text{test}$ to evaluate performance:
\begin{equation}
    Y_\text{test}=F(X_\text{test})\cdot W \longleftrightarrow \hat{Y}_\text{test}.
\end{equation}

We evaluate the performance of our quantum reservoir using two different metrics depending on the nature of the problem that we are dealing with. For discrete classification problems---those in Secs.\ \ref{sec:problemsGL} and\ \ref{sec:problemsGARCH}---the real-valued output produced by the reservoir is mapped to a class according to one or more thresholds. Performance is then quantified in terms of prediction accuracy, which we will denote by $A,$ and represents the percentage of correctly classified instances. On the other hand, the prediction of the degrees of freedom in a Student-\textit{t}  distribution, the problem in Sec.\ \ref{sec:problemsTS}, constitutes a continuous regression task. Here, we assess performance through the root mean square error (RMSE), which represents the average deviation between the predicted $\hat{y}_i$ and the target $y_i$ values:
\begin{equation} \label{eq:rmse}
\mathrm{RMSE} = \sqrt{\frac{1}{N} \sum_{i=1}^{N} \left( y_i - \hat{y}_i \right)^2 },
\end{equation}
where $N$ is the total number of predictions.

\section{Methods}

\subsection{QRC simulations}

We aim to numerically analyze QRC implemented in superconducting hardware described by Eq.\ \ref{eq:fullmodel}. However, any realistic analysis of a quantum information protocol must take into account the role of noise and its associated decoherence. To do so, we use the density matrix formulation of quantum dynamical evolution, which naturally allows the introduction of decoherence into the system\ \cite{ripoll2022quantum}. Under this formalism, the system is represented by a mixed state with an associated density matrix that we will denote by $\rho.$ The time evolution of this object is described by a master equation, which can be derived under the Markov approximation to be:
\begin{equation}\label{eq:mastereq}
    \dot{{\rho}} = -i [{H}, {\rho}] + \sum_{j=1}^2\bigg({C}_j {\rho} {C}_j^\dagger - \frac{1}{2} {C}_j^\dagger {C}_j {\rho} - \frac{1}{2} {\rho} {C}_j^\dagger {C}_j\bigg),
\end{equation}
where $H$ is the Hamiltonian associated with our hardware, see Eq.\ \ref{eq:fullmodel}, and ${C}_j$ are the collapse operators accounting for the decay of bosonic modes. These operators are given by
\begin{equation}
    {C}_j = \sqrt{\kappa_j}\, {a}_j,
\end{equation}
with $\kappa_j$ denoting the decay rate of mode $j$. Notice that we do not introduce pure dephasing coming from elastic processes, but decay due to inelastic processes. We have simulated Eq.\ \eqref{eq:mastereq} using the QuTiP library\ \cite{qutip5} with a bosonic cutoff in the occupancies of $n_c=5$. 

Throughout this work, we consider $\kappa_1 = \kappa_2 = 500 \,\mu\text{s}^{-1}$, which is quite reasonable for current experiments with superconducting circuits\ \cite{ihssen2025low}. Although this technology allows now for even longer coherence times, there is no need to push our implementation in that direction as QRC naturally benefits from a moderate amount of noise \cite{Domingo2023,Sannia2024dissipationas,Fry2023}. Hence, we deliberately operate in a regime in which dissipation is not very small. 

Now, we go into the specific details of the QRC algorithm. For each prediction task, the reservoir processes an entire dataset consisting of a sequence of $T$ real numbers, which are injected sequentially into it. At the beginning of each dataset, the reservoir is initialized in its ground state. Then, after the evolution associated with each element of the sequence, we measure the occupations of a subset of Fock states $\ket{i,j}$ with $0 \leq i,j \leq m$. This choice yields $(m+1)^2$ occupation probabilities, which we can interpret as the neurons of the reservoir. In the results presented in this work, we restrict to $m=2$, corresponding to a total of nine reservoir neurons.

Instead of using the raw time series of occupations directly, we construct the feature matrix $F(X)$ through a statistical pooling procedure. For each neuron $k$, we calculate a few temporal statistics over its trajectory, such as the mean, standard deviation, and correlation measures. These statistics summarize the dynamical response of the reservoir to the entire input sequence. Then, we concatenate the statistics of all neurons, which yields a single feature vector associated with that dataset. Finally, we stack the feature vectors corresponding to different datasets to produce the matrix $F(X)$ used in the linear readout. Notice that our QRC implementation does not require full quantum state tomography, but only the measurement of the occupation probabilities of the selected reservoir states. This is in contrast to other approaches that rely on full state reconstruction, which may be difficult to achieve in real quantum hardware.

 \subsection{Classical algorithms}

In order to evaluate the performance of QRC on the different tasks introduced in Sec.\ \ref{Sec:problems}, we compare its results with those obtained using classical benchmarking counterparts. We will use two main classical methods, a generalized likelihood ratio test (GLRT)\ \cite{kay1993fundamentals, lehmann2005testing} and a supervised classifier. The first of these methods is used for Gaussian/Laplace discrimination (Sec.\ \ref{sec:problemsGL}) and Student-t parameter inference (Sec.\ \ref{sec:problemsTS}). The second one is employed for the determination of GARCH volatility regimes (Sec.\ \ref{sec:problemsGARCH}). We briefly review each of those methods and how we apply them to the problems explained in Sec.\ \ref{Sec:problems}.

\subsection{Parameter estimation and uncertainty quantification.}

As we just said, GLRT is applied to the discrimination problem between Gaussian and Laplace distributions, Sec.\ \ref{sec:problemsGL}. For each input sequence, the parameters of the Normal and Laplacian distributions are estimated via maximum likelihood, and the sequence is assigned to the distribution for which the log-likelihood is larger. Specifically, given a sample $\{x_i\}_{i=1}^n$, we compare maximized log-likelihoods
\begin{equation}
\widehat{m} \;=\; \arg\max_{m\in\mathcal{M}} \; \max_{\theta_m}\; \sum_{i=1}^n \ell(x_i;\theta_m).
\end{equation}
Note that the log-likelihood functions are  $\ell_{\mathcal{M}}= \log(f_{\mathcal{M}})$, where the functions $f_{{\mathcal{M}}}$ are given in Eq.\ \eqref{eq:normal} and Eq.\ \ref{eq:laplace} for  $\mathcal{M}=\left\{\mathcal{N},\mathcal{L}\right\},$ respectively.  The standard  maximum likelihood estimators are $\hat\mu_{\mathcal{N}}=\frac{1}{n}\sum_i x_i$, $\hat\sigma^2_{\mathcal{N}}=\frac{1}{n}\sum_i (x_i-\hat\mu_{\mathcal{N}})^2$ for Gaussian and $\hat\mu_{\mathcal{L}}=\mathrm{median}(x_i)$, $\hat b_{\mathcal{L}}=\frac{1}{n}\sum_i |x_i-\hat\mu_{\mathcal{L}}|$ for Laplace distribution, respectively.

For the Student-$t$ degrees-of-freedom estimation task in Sec. \ref{sec:problemsTS}, the classical algorithm reference is also maximum-likelihood estimation of $\nu$ 
\begin{equation}
\hat\nu=\arg\max_{\nu}\sum_{i=1}^n \log f_t(x_i;0,1,\nu),
\end{equation}
where $f_t$ is given by the probability density function of the Student-t distribution given in Eq.\ \ref{eq:tstudent}. We report performance in terms of $1/\nu$  to emphasize the low-$\nu$ regime.
 
The classical method used as a reference for GARCH problem in Sec.\ \ref{sec:problemsGARCH} consists of computing the same features as in our QRC approach, but over the raw data $x_t$ rather than the neuron populations of the reservoir. These features are used to train a supervised classifier on a training set, which is then evaluated on a separate test set. This classifier is equivalent to the last readout layer of the QRC. Thus, this approach is based on the prediction given by the matrix $W$ in Eq.\ \ref{eq:pseudo}, but using the $X_{training}$ as the raw data of the input signal. This is one of the standard classical methods for GARCH regime classification in time-series analysis and volatility modeling \cite{BOLLERSLEV1986307, bollerslev1992quasi, Tsay2010, Brooks2019}.

\section{Results}

We now analyze the efficiency of QRC and classical methods when solving the problems of statistical inference explained in Sec.\ \ref{Sec:problems}. The reservoir dynamics correspond to those of our superconducting system in Fig.\ \ref{fig:circuito_esquema}, which, as we derived previously, is described by the Bose-Hubbard model Eq.\ \ref{eq:fullmodel}. We fix the system parameters to $\omega_1 = 10\,\mathrm{GHz}$, $\omega_2 = 9\,\mathrm{GHz}$, $g_{12} = 400\,\mathrm{MHz}$, and $U = 600\,\mathrm{MHz}$. The value of $\varepsilon_0$ varies slightly across tasks and will be reported in each of them. 

In each of the problems tackled we characterize the performance of the prediction in terms of a quantity $X$, and fit the results to laws of the form $X(T)$, where $T$ is the number of data points fed to the algorithm. When displaying the results, the reported markers correspond to the mean performance over independent test realizations, while error bars denote $\pm 1\sigma$ (one standard deviation) around the mean.  Fitting the scaling laws $X(T)$ allows for a qualitative understanding of the performance of QRC compared with classical methods. A detailed explanation of the fitting procedure that we have followed can be found in the Supplementary Material.  

For the two classification tasks (Normal vs.\ Laplace and GARCH-band assignment), we use the accuracy of the prediction $X=A$ as a performance characterization. This accuracy $A$  is defined as the percentage of successful predictions when solving many different problems with equal-length $T$ data inputs. We then fit the accuracy of the predictions as a function of $T$ to obtain a scaling law. We consider two main forms for this scaling law
\begin{align}
A(T) &= A_\infty - c\,e^{-kT^{p}}, \label{eq:acc-exp}\\
A(T) &= A_\infty - c\,T^{-p} \label{eq:acc-power}.
\end{align}
For the Student-\textit{t} regression task we use the prediction error $X=\mathrm{RMSE}$, see Eq.\ \ref{eq:rmse},  on the target $1/\nu$ . Specifically, we test which of the following scaling laws fit our data better:
\begin{align}
\mathrm{RMSE}(T) &= c\,T^{-p}, \label{eq:rmse_power}\\
\mathrm{RMSE}(T) &= c\,(\ln T)^{-p}\qquad (T>1),\label{eq:rmseII}\\
\mathrm{RMSE}(T) &= r_\infty + c\,e^{-kT^{p}},\label{eq:rmseIII}
\end{align}
where $T$ is the number of data points fed into the algorithm, as before.

\begin{figure}[htbp]
    \centering

    \begin{subfigure}{\linewidth}
        \centering
        \begin{tikzpicture}
            \node[inner sep=0] (img) {\includegraphics[width=\linewidth]{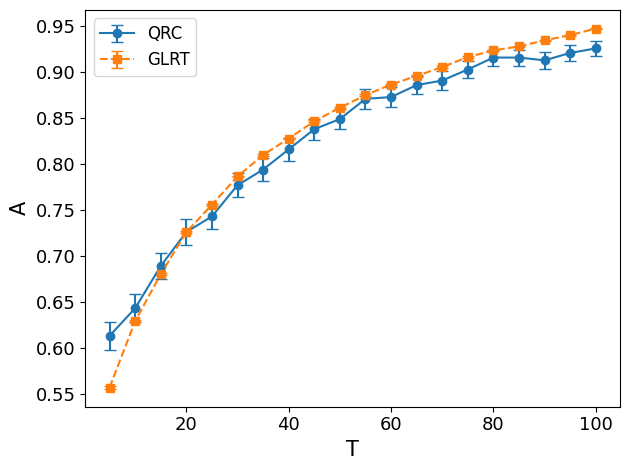}};
            \node[anchor=north west] at ([xshift=2mm,yshift=8mm]img.north west) {\large{$(a)$}};
        \end{tikzpicture}
        \label{fig:nvsl_a}
    \end{subfigure}

    \vspace{1em}

    \begin{subfigure}{\linewidth}
        \centering
        \begin{tikzpicture}
            \node[inner sep=0] (img) {\includegraphics[width=\linewidth]{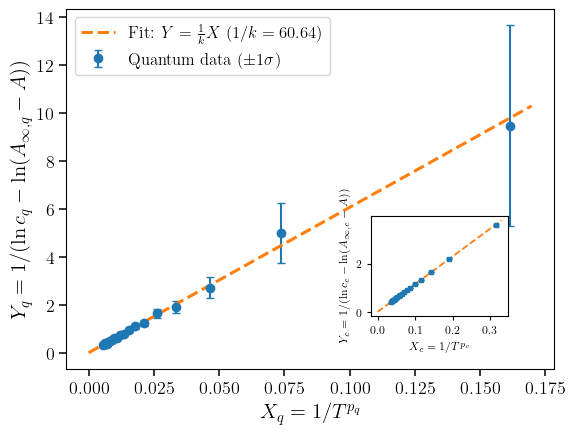}};
            \node[anchor=north west] at ([xshift=2mm,yshift=8mm]img.north west) {\large{$(b)$}};
        \end{tikzpicture}
        \label{fig:nvsl_b}
    \end{subfigure}

    \caption{Normal vs.\ Laplace discrimination. (a) Prediction accuracy of QRC and a generalized likelihood ratio test versus the number of data points per sample $T$. (b) Accuracy scaling of QRC with input length $T$ fitted by stretched-exponential laws $A(T)=A_{\infty}-c\,e^{-kT^{p}}$, together with the corresponding linearized representation (inset: classical case).
    For this task, the driving amplitude is fixed to $\epsilon_0 = 3.8\,\mathrm{GHz}$.}
    \label{fig:nvsl}
\end{figure}

\begin{figure}[h]
    \centering

    \begin{subfigure}{\linewidth}
        \centering
        \begin{tikzpicture}
            \node[inner sep=0] (img) {\includegraphics[width=\linewidth]{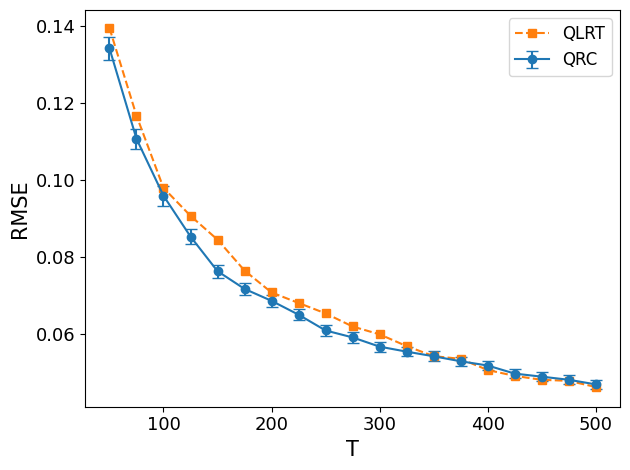}};
            \node[anchor=north west]
                at ([xshift=2mm,yshift=8mm]img.north west) {\large{$(a)$}};
        \end{tikzpicture}
        \label{fig:tstudent_a}
    \end{subfigure}

    \vspace{0.6em}

    \begin{subfigure}{\linewidth}
        \centering
        \begin{tikzpicture}
            \node[inner sep=0] (img) {\includegraphics[width=\linewidth]{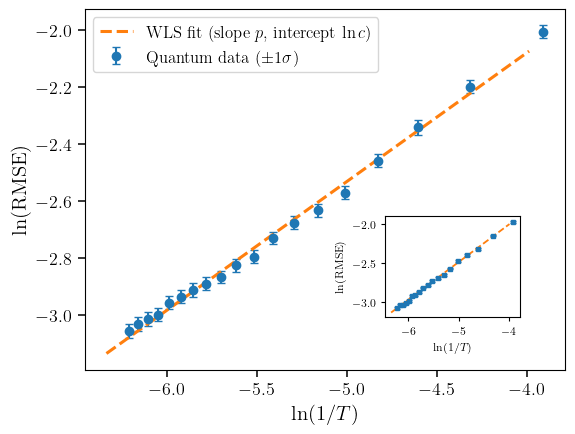}};
            \node[anchor=north west]
                at ([xshift=2mm,yshift=8mm]img.north west) {\large{$(b)$}};
        \end{tikzpicture}
        \label{fig:tstudent_b}
    \end{subfigure}
    \caption{Student-\textit{t} degrees-of-freedom prediction. (a) Test-set RMSE of the $1/\nu$ estimator versus sequence length $T$, comparing the QRC readout with a classical maximum-likelihood estimator.  Both methods improve monotonically with $T$, with QRC achieving lower RMSE and the gap narrowing at large $T$.  (b)  Scaling analysis of data in panel (a), which are fitted to power-law decays $RMSE(T)=c\,T^{-p}$, together with the corresponding linearized representation (inset: classical case). 
    For this task, the driving amplitude is fixed to $\epsilon_0 = 1\,\mathrm{GHz}$.
    }\label{fig:tstudent}
\end{figure}

\subsection{Normal vs.\ Laplace Classification}
The first task probes whether the reservoir can identify distributional \emph{shape} under random translations and re-scalings. The QRC and GLRT accuracies for this task versus the length of the data sequence $T$ are shown in Fig.~\ref{fig:nvsl}(a). We observe that QRC is competitive against the generalized likelihood ratio test (GLRT) across the explored range of sample sizes, and it is particularly strong in the small-$T$ regime. That is, QRC attains a significantly higher accuracy than GLRT in the case of short data sequences, with a separation comparable to (or larger than) the displayed uncertainty. As $T$ increases, the two methods rapidly converge. At the largest sample sizes, GLRT seems to perform better. However, the confidence intervals for both cases still overlap substantially for most points. The scaling analysis in Fig.~\ref{fig:nvsl}(b) indicates that both datasets are best summarized by stretched-exponential approaches to an asymptote, see Eq.\ \ref{eq:acc-exp}. The best-fit parameters are $A_{\infty,c}=1.000\pm0.005$ with $(c_c,k_c,p_c)=(0.58\pm0.02,0.086\pm0.008,0.72\pm0.03)$ for GLRT, and $A_{\infty,q}=0.943\pm0.007$ with $(c_q,k_q,p_q)=(0.37\pm0.02,0.016\pm0.005,1.14\pm0.08)$ for QRC, with $\chi^2_{\rm red}=2.85$ (classical) and $\chi^2_{\rm red}=0.17$ (quantum). These fits are consistent with the qualitative behaviour observed in Fig.\ \ref{fig:nvsl}(a): both approaches improve quickly with $T$, and the classical method outperforms QRC at large $T,$ as indicated by $A_{\infty,c}>A_{\infty,q}.$

\subsection{Prediction of Student-\textit{t} Degrees of Freedom}

We next consider a regression setting in which the target is the inverse of the degrees of freedom of a Student-t distribution, $1/\nu.$ As we have previously commented, we use this quantity to enhance sensitivity to tail heaviness. Figure~\ref{fig:tstudent} contains the RMSE of the prediction as a function of the sequence length $T$ for the quantum and classical cases. In contrast to the previous task, QRC shows a systematic advantage over the likelihood-based protocol across a broad interval of $T.$ From short to intermediate series lengths, the QRC curve lies below the classical one, and the separation is most pronounced around the intermediate-$T$ range, where the error bars do not overlap. For the longest series lengths, both methods approach a similar error floor and the difference becomes comparable to the error bars. The scaling fits shown in Fig.~\ref{fig:tstudent}(b) select a power-law decay given by Eq.\ \ref{eq:rmse_power} for both approaches, with exponents $p_c=0.483\pm0.005$ and $p_q=0.451\pm0.008$ and prefactors $c_c=0.93\pm0.02$ and $c_q=0.76\pm0.03$ (with $\chi^2_{\rm red}=1.07$ and $0.85$, respectively). The similar exponents indicate comparable asymptotic decay rates with $T$, whereas the smaller prefactor obtained for QRC compactly captures the lower RMSE observed throughout most of the explored range. Furthermore, we obtain an exponent close to $p=0.5.$  This is the exponent expected when taking into account the Central Limit Theorem. Indeed, for any regular estimator built from $T$ effectively independent samples, that theorem implies $\sqrt{T}\,(\hat\theta-\theta)\Rightarrow\mathcal N(0,\sigma_\theta^2)$, and hence $\mathrm{RMSE}(\hat\theta)\sim \sigma_\theta/\sqrt{T}$, i.e.\ a $T^{-1/2}$ decay.
\begin{figure}[htbp]
    \centering
    \begin{subfigure}{\linewidth}
        \centering
        \begin{tikzpicture}
            \node[inner sep=0] (img) {\includegraphics[width=\linewidth]{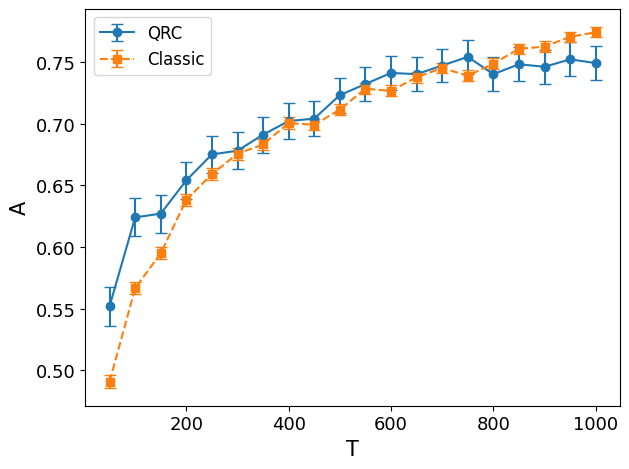}};
            \node[anchor=north west]
                at ([xshift=2mm,yshift=8mm]img.north west) {\large{$(a)$}};
        \end{tikzpicture}
        \label{fig:garch_a}
    \end{subfigure}
    \vspace{0.6em}
    \begin{subfigure}{\linewidth}
        \centering
        \begin{tikzpicture}
            \node[inner sep=0] (img) {\includegraphics[width=\linewidth]{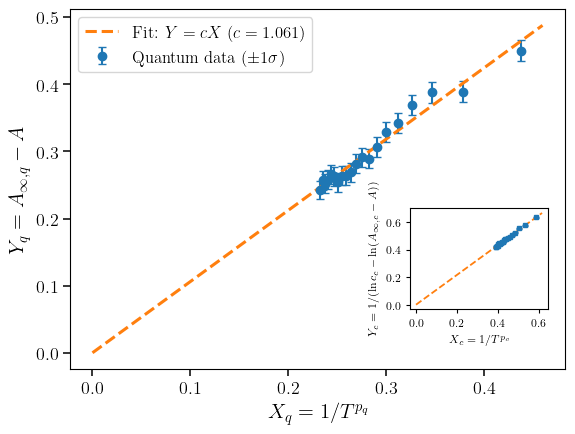}};
            \node[anchor=north west]
                at ([xshift=2mm,yshift=8mm]img.north west) {\large{$(b)$}};
        \end{tikzpicture}
        \label{fig:garch_b}
    \end{subfigure}

    \caption{GARCH(1,1) volatility-regime classification. (a) Test accuracy versus input length $T$ for QRC and a classical feature-based classifier. (b) Scaling analysis of the accuracy versus $T$ using the fitted scaling laws for $A(T).$
    The classical method is best described by a stretched-exponential approach to a plateau, $\hat A_c(T)=A_{\infty,c}-c_c e^{-k_c T^{p_c}}$, whereas the QRC data favor a power-law approach, $\hat A_q(T)=A_{\infty,q}-c_q T^{-p_q}$. 
    For this task, the driving amplitude is fixed to $\epsilon_0 = 1.9\,\mathrm{GHz}$.}
    \label{fig:garch}
\end{figure}

\subsection{GARCH Band Classification}
Finally, we address the correlated setting of $\mathrm{GARCH}(1,1)$ sequences, where the goal is to assign each time series to one of three persistence bands. Figure~\ref{fig:garch} shows that QRC provides a benefit for short and moderately short time series: in the low-$T$ regime, QRC achieves higher accuracy than the feature-based classical classifier trained directly on the raw $x_t$, and the improvement is larger than (or comparable to) the corresponding confidence intervals. As $T$ increases, the two approaches become nearly indistinguishable within uncertainty, and at the largest $T$ values the classical curve can slightly exceed QRC, albeit with overlapping error bars. 

\begin{table*}[htbp]
\centering
\renewcommand{\arraystretch}{1.35}
\setlength{\tabcolsep}{4pt}

\begin{tabular}{@{} l l l l l l r @{}}
\toprule
\textbf{Task} &
\textbf{Selected law} &
\textbf{$A_\infty/r_\infty$ ($\pm\sigma$)} &
\textbf{$c$ ($\pm\sigma$)} &
\textbf{$k$ ($\pm\sigma$)} &
\textbf{$p$ ($\pm\sigma$)} &
\textbf{$\chi^2_{\rm red}$} \\
\midrule
\shortstack[l]{GARCH\\classic} &
\eqref{eq:acc-exp} &
$1.0000\pm0.0013$ &
$2.47\pm0.08$ &
$0.926 \pm 0.014$ &
$0.136\pm0.008$ &
1.40 \\
\midrule
\shortstack[l]{GARCH\\quantum} &
\eqref{eq:acc-power} &
$1.00\pm0.13$ &
$1.01\pm0.06$ &
\text{--} &
$0.21\pm0.08$ &
0.56 \\
\midrule
\shortstack[l]{Normal vs\\Laplace classic} &
\eqref{eq:acc-exp} &
$1.000\pm0.005$ &
$0.58\pm0.02$ &
$0.086\pm0.008$ &
$0.72\pm0.03$ &
2.85 \\
\midrule
\shortstack[l]{Normal vs\\Laplace quantum} &
\eqref{eq:acc-exp} &
$0.943\pm0.007$ &
$0.37\pm0.02$ &
$0.016\pm0.005$ &
$1.14\pm0.08$ &
0.17 \\
\midrule
\shortstack[l]{$t$-Student $\nu$\\classic} &
\eqref{eq:rmse_power} &
\text{--} &
$0.93\pm0.02$ &
\text{--} &
$0.483\pm0.005$ &
1.07 \\
\midrule
\shortstack[l]{$t$-Student $\nu$\\quantum} &
\eqref{eq:rmse_power} &
\text{--} &
$0.76\pm0.03$ &
\text{--} &
$0.451\pm0.008$ &
0.85 \\
\bottomrule
\end{tabular}
\caption{Summary of the selected scaling laws and the corresponding fitted parameters for each task. 
The second column contains the selected law, the one with the closest to one $\chi^2_r,$ from the ansatzs Eqs.\ \eqref{eq:acc-exp}-\eqref{eq:rmseIII}. The other columns indicate the estimation of parameters  $\pm 1\sigma$ uncertainties, together with the reduced chi-squared of the fit in the last column, $\chi^2_{\rm red}.$ }
\label{tab:scaling_summary}
\end{table*}

The scaling fits in Fig.~\ref{fig:garch} favor different effective laws for the two approaches. For the classical baseline, the best description is given by Eq. \eqref{eq:acc-exp} with $A_{\infty,c}=1.0000\pm0.0013$, $(c_c,k_c,p_c)=(2.47\pm0.08,0.926\pm0.014,\,0.136\pm0.008)$ and $\chi^2_{\rm red}=1.40$, while for QRC the preferred ansatz is Eq. \eqref{eq:acc-power}, with $A_{\infty,q}=1.00\pm0.13$, $p_q=0.21\pm0.08$, $c_q=1.01\pm0.06$ and $\chi^2_{\rm red}=0.56$. At the phenomenological level, these fits reflect the same qualitative message as the raw curves: QRC extracts useful regime information already from relatively short correlated sequences, whereas for longer time series the classical method exhibits better accuracy in its predictions.  We further observe a peculiar swing of the quantum data points around the  theoretical fitted law in Fig.\ \ref{fig:garch}(b), which does not occur in the other problem treated above. This correlation in the accuracy of the prediction for different time series may be caused by the 
intrinsic time-correlations exhibited by GARCH models.

\subsection{Performance overview}

Table~\ref{tab:scaling_summary} summarizes the scaling-law fits obtained for each task. We report there the selected functional form from the non-linear analysis together with the associated parameter estimates and complementary goodness-of-fit diagnostics. In particular, we provide the asymptotic performance level ($A_\infty$ for accuracies, or $r_\infty$ for RMSE when applicable) together with its uncertainty, the fitted coefficients $(c,k,p)$, and the reduced chi-squared values computed from the  non-linear weighted least-squares fits, see Supplementary Material. 

We have seen that QRC improves over classical algorithms at short $T,$ where decisions must be made with limited data.  In contrast, classical methods marginally outperform QRC when the amount of data is large. This is reflected in the parameters of the fitted laws. Notice that the parameters that control the efficiency as $T\rightarrow \infty$ are consistent with asymptotic accuracies close to unity in most cases (not in the Normal/Laplace discrimination). This is not trivial, as there is no general theoretical guarantee that QRC should saturate to unit accuracy in the infinite-data limit. This is the case at least in the GARCH problem, as time correlations must be taken into account for correct parameter estimation. The good performance of QRC in the GARCH problem is likely to be caused by time correlation being smaller than the noise-induced decay of the quantum state. As if it were the opposite, decoherence would produce a loss of memory likely to be incompatible with a full GARCH parameter estimation at infinite time.

Once we have seen that the efficiency of QRC can be as good as desired by increasing the sequence length, we can focus on the parameters that control the finite-time prediction results. There are two main parameters that control the efficiency in  this regime. The ones that set the long time scaling are those denoted by $p$, while those responsible for the short-$T$ limit are denoted by $c$ in Eqs.\ \ref{eq:acc-exp} to \ref{eq:rmseIII}. Our parameter estimates systematically indicate that, in all the problems treated,  QRC beats classical methods at short $T$ while having lower efficiency in the long-$T$ limit.

\section{Conclusions}

Our results indicate that QRC with Bose-Hubbard-type dynamics can be more efficient than classical algorithms solving problems in statistical inference in the regime of limited data. From the problems we analyzed, the GARCH problem---explained in Sec.\ \ref{sec:problemsGARCH}---is the most relevant one for financial forecasting. At the same time, it is  the most difficult problem to solve due to data correlations and heavy tails. For this specific computational task, we obtained the clearest advantage of QRC with respect to its classical counterpart in the regime of limited information.  This regime, where QRC outperforms the classical method, is particularly relevant. Indeed, one would like to have clear information about the GARCH volatility regime as soon as possible in order to optimize portfolio investments.  We have obtained similar results for the other statistical inference tasks analyzed here, distribution discrimination and Student-t parameter estimation, for which QRC provided a quantum advantage in the limit of scarce information. 

The above advantages have been obtained using artificially constrained reservoir dynamics. Indeed, our numerical computations have been performed using a cutoff in the bosonic occupancies in order to make the numerical simulation feasible. Taking into account the rather good efficiency of our \emph{cut-off} QRC-based learning, we  strongly believe that it is worth implementing the same algorithm on real superconducting hardware.  Running our numerical algorithm on real quantum hardware will make it possible to access reservoirs with a much larger number of neurons. Shuch an increase in the number of neurons will likely to induce a change in the underlying reservoir dynamic. Indeed, we have discussed how the Bose-Hubbard model Eq.\ \ref{eq:fullmodel} may not capture the stronger non-linear evolution in the case of large bosonic cut-offs, as a full rotor model is likely to be needed. 

In summary, we have laid the foundations for constructing analog superconducting circuits that can be used to implement QRC for statistical inference. This avenue of research offers many ingredients that can be added to our basic algorithm and to the experimental setup, potentially leading to improvements in QRC performance. As discussed in the previous paragraph, a larger Hilbert space would likely benefit QRC. On top of that, one could envision adding more nonlinearities to the reservoir dynamics, either by operating the Josephson junctions in a different regime or by incorporating a few additional superconducting islands with all-to-all couplings. Notice that all-to-all coupling naturally occurs in systems of capacitively connected superconducting islands due to the long-range character of the Coulomb interaction. All these directions are within the reach of experimentally available superconducting systems, which makes this an interesting topic to explore further.

\section{Acknowledges}

This work is part of the European Union NextGeneration EU/PRTR project Consolidación Investigadora CNS2022-136025. M. P. acknowledges further support through grant no. PID2024-156340NB-I00 funded by Ministerio de Ciencia, Innovación y Universidades/Agencia Estatal de Investigación (MICIU/AEI/10.13039/501100011033) and the European Regional Development Fund (ERDF). We gratefully acknowledge funding by the University of Warwick’s International Partnership Fund 2024. The numerical computations were performed in the facilities of Supercomputación Castilla y León (SCAYLE).

\bibliographystyle{unsrt}
\bibliography{references}

\clearpage
\onecolumngrid

\renewcommand{\thesection}{S\arabic{section}}
\setcounter{section}{0}


\begin{center}
{\large \bfseries Supplemental Material for:  Quantum Reservoir Computing for Statistical inference in a analog superconducting circuit} \\[1em]
\end{center}

\section{Fitting procedure}

We explain the procedure used to fit data to the theoretical laws in each of the tasks described in the main body of the manuscript. The fits are performed in a weighted manner, using one-standard-deviation uncertainties to normalize the residuals. In practice, we followed two complementary procedures.  First, we carried out a fit of the data to each of the candidate laws described above, Eqs.\ \ref{eq:acc-exp}, \ref{eq:acc-power} for classification tasks and Eqs.\ \ref{eq:rmse_power}-\ref{eq:rmseIII} for parameter estimation.  We then select the candidate whose reduced chi-squared $\chi^2_{\rm red}$ is closest to unity.  Second, to place the $\chi^2_{\rm red}$-based comparison on firmer statistical footing, we repeated the analysis using linearized weighted regressions. Note that the reduced chi-squared diagnostic is strictly justified for (weighted) linear least-squares models \cite{andrae2010dosdontsreducedchisquared}.  Both procedures led to consistent model and parameter estimates.

Let us explain in more depth the procedure for the linearized approach. We used a grid of values for a subset of the parameters. For each point of the grid, we performed a linear fit to extract the values of the other parameters and the $\chi_{\rm red}^2$ of the fit. We then select the point of the grid that yields a $\chi_{\rm red}^2$ closest to unity and report the value of the scanned parameters together with the uncertainties determined from the grid resolution. For instance, for the power law $A(T)=A_\infty-c\,T^{-p}$, we scan $A_\infty$ and fit $\ln(A_\infty-A)$ versus $\ln T$ to extract $(\ln c,p)$ by weighted linear regression. For the stretched-exponential form $A(T)=A_\infty-c\,e^{-kT^p}$, we scan $(A_\infty,p)$ and fit $\ln(A_\infty-A)$ versus $T^p$ to extract $(\ln c,k)$. For the Student-\textit{t} regression task, we fit the error curve $\mathrm{RMSE}(T)$ for the target $1/\nu$. The power and log-power candidates are handled via weighted linear regression after taking logarithms, i.e.\ $\ln \mathrm{RMSE}$ versus $\ln T$ and $\ln \mathrm{RMSE}$ versus $\ln\ln T$, respectively (yielding $(\ln c,p)$), whereas the exponential-with-floor candidate $\mathrm{RMSE}(T)=r_\infty+c\,e^{-kT^p}$ is evaluated by scanning $(r_\infty,p)$ and fitting $\ln(\mathrm{RMSE}-r_\infty)$ versus $T^p$ to obtain $(\ln c,k)$. 

When fitting scaling laws to performance curves in the non-linear approach, we estimate parameters by weighted non-linear least squares (WLS), using residuals normalized by the reported $\sigma$ uncertainties of each data point. Specifically, for observations $y_i$ at $T_i$ with error bars $\sigma_i$, we minimize
\begin{equation}
\chi^2(\boldsymbol\theta)=\sum_i \left[\frac{y_i-f(T_i;\boldsymbol\theta)}{\sigma_i}\right]^2,
\end{equation}
and report parameter uncertainties as $\pm1\sigma$ from the Gauss--Newton (local linearization) approximation to the covariance at the optimum $\hat{\boldsymbol\theta}$,
\begin{equation}
\widehat{\mathrm{Cov}}(\hat{\boldsymbol\theta}) \;\approx\; \chi^2_{\rm red}\,\big(J^\top J\big)^{-1},
\end{equation}
where $J$ is the Jacobian of the normalized residual vector with respect to $\boldsymbol\theta$ evaluated at $\hat{\boldsymbol\theta}$ and $\chi^2_{\rm red}=\chi^2/{\rm dof}$.
We report the results of these non-linear fits, as they are consistent with those obtained from the linearized approach described in the previous paragraph. 


\end{document}